\input harvmac.tex
\Title{\vbox{\baselineskip14pt\hbox{HUTP-98/A026}\hbox{hep-th/9805213}}}
{Minimal Cycles, Black Holes and  QFT's}
\bigskip\vskip2ex
\centerline{Cumrun Vafa}
\vskip2ex
\centerline{\it  Lyman Laboratory of Physics, Harvard
University}
\centerline{\it Cambridge, MA 02138, USA}
\vskip .3in
We review some aspects of
minimal cycles in string compactifications and
their role in constructing new critical theories in
six and lower dimensions as well as in accounting for black hole entropy.
(Based on a talk presented at the Salam Memorial Meeting, the Abdus Salam
International Center for Theoretical Physics, Fall 1997)

\Date{\it {May 1998}}
\newsec{Some Reflections on Abdus Salam}
Before I start with the scientific aspect of this note, I would like
to mention a few words about my impressions of Abdus Salam and his
contributions to physics and to the advancement of science
in developing countries.

Science and scientific tradition has a long history. At different
times different countries have led the cause of advancing
human knowledge.  Many of the countries that today are called
``developing countries'' or ``third world countries'' have
at an earlier time been at the forefront of knowledge and have
made important contributions to science.  Indeed the present scientific
tradition is based on the foundations set by these earlier advances.
 The very fact that science has developed in this way also
clearly demonstrates that science does not belong to a restricted
group of people.  It belongs to all humanity.  Thus all humanity
and in particular scientists share a responsibility of making pursuit of
science as much of a global, borderless endeavor, as they can.

In the present century if one is to name a single person who
has contributed most to the development of science in the developing
countries the natural choice would be Abdus Salam.  With the creation
of the International Center for Theoretical Physics he has contributed
tremendously to making modern science accessible to a much
larger audience of humanity.
 Through this center and through
numerous visits to developing countries he encouraged and challenged
the citizens of these countries to take up the cause of advancing
science.   In fact the ICTP has gone far beyond a center for developing
countries, and has become a center which attracts first rate scientists from
all over the world and has become a regular sight for important
conferences in physics and in mathematics.

The creation of the ICTP was a rather difficult task.  He always
had a tough task of convincing authorities for the need of having such
a center and the need for improvement of science in the developing countries.
The irony of this is that he had an easier time convincing authorities
of developed countries in contributing in this direction than those of the
developing countries.  To the policy makers of the developing countries he
emphasized an important fact:
It may appear that advancing modern abstract science in developing
countries is a futile task, when one can import any advanced technological
product.  However the progress of a country will ultimately come
from having a strong scientific base, even though it may appear in the
short run to be far from any concrete application.

Even though this aspect of his life has had such a crucial
impact in the lives of so many scientists, his contributions to theoretical
physics are no less important.   In particular
his contributions to particle physics resulted in his discovery of what we
now call the Salam-Weinberg standard model.
Since this aspect of his life and his scientific
contributions in general are already very well known
I will not write any further on this point.

\newsec{Basic Idea}
Perhaps the most important aspect of the recent revolution
in string theory has been the appreciation of the role that
extended objects play in physics\foot{I will not include any references in this
short note.  The interested reader can consult the many good review
articles already available.}.  These objects can come in various
dimensionalities, and are typically viewed as solitonic degrees of freedom.
If they have $p$ spatial dimensions, they are called $p$-branes (extending
the terminology from membrane where $p=2$).  The existence of such
states, together with the
properties of the internal geometry $M$ where the string is compactified
upon leads to many interesting and novel physical phenomena.  In particular
the $p$-brane can wrap around a $q$-dimensional cycle $C\subset M$
of the internal
manifold leading to a $p-q$ dimensional brane in the uncompactified
space.  The basic idea is to investigate what kinds of $C$'s there are
and what physical consequences they lead to.  There are basically
two classes of cycles that we will consider; the cycles $C$ that
can shrink to zero size, and are thus called the {\it vanishing cycles},
and cycles that are typically big and cannot shrink beyond a certain minimum
size.   The type that can be shrunk are typically ``rigid'', which means
that they do not have any moduli associated with them.  The second type,
typically come in a family, i.e. there is a moduli space associated
with them (what this means is that if we look at minimal cycles
in a given homology class there is more than one, and they can
be parameterized by some space called the moduli space).
We shall see that the first type will lead to interesting quantum
field theory questions whereas the latter type have bearing
on questions involving black holes and their entropy.

\newsec{Vanishing Cycles and QFT Interpretations}
Consider a string compactification with a $p$-dimensional vanishing
cycle.  If we wrap a $p$-brane around such a cycle, the resulting
state will be a 0-brane, i.e. a particle, from the viewpoint of the
uncompactified spacetime.  Moreover the mass of this state is given by
$$M=T\cdot V(C)$$
where $T$ denotes the tension on the worldvolume of the $p$-brane
and $V(C)$ denotes the volume of $C$.  The assumption that the cycle
is vanishing means that $V(C)\rightarrow 0$, which thus implies
that we end up with massless particles.  Depending on various
cases and the different values of $p$ these massless particles
will have different implications.  For example if $p=2$ and these
vanishing 2 spheres arise in a type IIA string
compactified on $K3$ down to six dimensions, this
gives rise to massless vector multiplet charged under a $U(1)$,
which thus naturally enhances the gauge symmetry from $U(1)\rightarrow
SU(2)$.  If there are more than one vanishing 2-cycle and they
intersect one another according to the Dynkin diagrams of the $A-D-E$
groups, the resulting physics is an enhanced A-D-E gauge symmetry.
If we further compactify down to four dimensions and we have a 2-dimensional
locus with $A_{k-1}$ singularity and another locus with $A_{p-1}$ singularity
and they meet at a point, the mixed wrapped 2-cycles will now lead
to $({\bf k,p})$ bi-fundamental matter of $SU(k)\times SU(p)$ gauge group
in four dimensions.
In fact by arranging various intersecting singularities
we can {\it engineer} and study the properties of a large class of quantum
field theories in this way.  Many non-perturbative questions of quantum
field theories get translated to perturbative string questions.

Sometimes it may happen that a $q$-dimensional cycle shrinks,
but the theory in question has no $q$-branes.  In such a case the
lightest states comes from a $q+r$ dimensional brane wrapped around it,
with the smallest available $r$, which typically is $r=1$.  In this case
we would end up with a tensionless string.   An example of this
is type IIB compactification on $K3$ with vanishing 2-cycle, in which
case we have no 2-branes, but one can consider wrapping the available 3-brane
of type IIB around the 2-cycle and this results in a tensionless string
in 6 dimension.  Such cases are a source of new phenomena in quantum field
theories, and are believed to be resulting in new non-trivial conformal
quantum field theories.  The example just mentioned would be a non-trivial
six dimensional critical quantum field theory.  Similar examples
can be constructed also upon compactification to lower dimensions.

\newsec{Black Hole applications}

Now we consider the case where the cycle $C$ cannot be made too small.
These cases are also correlated with the fact that the cycle $C$
can be deformed inside the compactification geometry, while
preserving its minimal area.
Let us consider $K3$ compactification
of type IIA with a fixed volume.  Consider a genus $g$
Riemann surface $\Sigma$ in it with minimal area (which implies
that it is embedded
holomorphically
with some choice of complex structure on $K3$).  In this case one can
show that the volume of $C$ is always bigger than
$$V(C)\geq {\rm const}. \sqrt g$$
Let us now consider wrapping a 2-cycle in type IIA theory compactified
on $K3$ about a genus $g$ Riemann surface $\Sigma$.  The resulting state will
be a 0-brane (i.e. a particle) in 6 dimension with mass proportional
to $V(C)$.  Now let us consider the case where $g$ is large.   In this case,
using the above bound, we find that its mass becomes huge (of the order of
$\sqrt g$).  The resulting state can be viewed as a very massive charged
particle in six dimensions.  In fact it looks like a macroscopic black hole.
One could then ask about comparisons between the area of the horizon
and the number of such cycles $C$ and see if there is any relation, as would
be expected based on Bekenstein-Hawking entropy formula.
In this particular case the corresponding black hole solution
has a singular horizon and so no exact reliable
area can be extracted from it.  Instead if we consider type IIB
compactification on $K3\times S^1$ and wrap a 3-brane around a Riemann
surface $\Sigma$ of genus $g$ in $K3$ times the $S^1$ and in addition consider
a state with some momentum $p$ along $S^1$ the corresponding
black hole will have a non-singular horizon and one can compare
the area of the horizon $A$ with the number
of microscopic degrees of freedom of the 3-brane in that state.

The way to compute the microscopic degrees of freedom of this
3-brane is to first consider the effective 1+1 dimensional theory
one gets after wrapping the 3-brane around the Riemann surface
$\Sigma$.  The number of degrees of freedom in the 1+1
dimensional theory will be related to number of ways the
cycle $C$ (together with a choice of gauge field on it) can be
deformed in $K3$.  This gives a theory in 1+1 dimension with an
effective $c=6g$ degrees of freedom ($4g$ coming from bosonic modes
and $2g$ from fermionic modes).  Then one looks at the number of
states in this 1+1 dimensional theory on a circle with momentum $p$.
This goes as
$$N\sim {\rm exp}[2\pi \sqrt{ p c/6}]={\rm exp}[2\pi\sqrt{ p g}]$$
which agrees with the predicted entropy
$$S={1\over 4}A=2\pi\sqrt{ p g}.$$

Similarly one can extend this analysis and construct
other classes of black holes in 5 and 4 dimensions and compare
the microscopic entropy with the predicted macroscopic entropy.

\vglue 2cm
This research was supported in part by NSF grant PHY-92-18167.

\end